# On a non-linear sigma model of knotted relaxed states far from thermodynamic equilibrium in plasma physics and beyond

A. Di Vita, DICCA, Università di Genova, Via Montallegro 1, 16145 Genova, Italy

*Abstract*. We show that a Faddeev-Niemi non-linear sigma model describes in the long wavelength limit a wide class of steady-state, knotted physical systems far from thermodynamic equilibrium which are stable against perturbations of temperature and interact weakly with the external world. In these systems temperature gradients are negligible, inertial effects are negligible in comparison with diffusion effects, entropy is mainly produced through Joule and/or viscous heating, the macroscopic state is described by specifying a unit vector **n**(**x**) at each point, and the Gauss linking number of **n**(**x**) is lower than a threshold. In fluids and plasmas, the model describes filamentary structures which adjust themselves in order to offer minimum resistance to the medium embedding them and to the electric currents (if any) flowing across them; in the latter case, Gauss linking number is related to magnetic helicity. Both **n**(**x**) and the relative velocity of the filament with respect to the medium are approximately Double Beltrami vector fields. We derive a stability criterion for a double helix. Moreover, a similar discussion describes the recently discovered 'writing' process of skyrmions in a magnetic film with the help of a beam of polarised electrons. We derive a lower bound on the value of beam current required to 'write' a skyrmion.

**1 The problem.**  Spontaneous evolution of physical systems towards relatively long-lived, highly organized, steady ($\partial / \partial t = 0$) states is usually referred to as 'relaxation'. (Here and in the following we are going to refer to a generic field **a**(**x**) in steady state as to a mean, fluctuation-averaged quantity). Relaxed configurations may be kept far from thermodynamical equilibrium by suitable boundary conditions (e.g. applied electric voltage, or constant incoming flows across the boundary) [1]; accordingly, the usual equilibrium thermodynamics does not apply. Generally speaking, various dissipative (i.e., entropy-raising) mechanisms may affect the geometrical structure of relaxed states [2]. For example, the velocity field **v** which solves Navier-Stokes equation in a steady-state fluid with Reynolds number $Re \ll 1$ satisfies Kortweg-Helmholz' variational principle, i.e. minimization of total viscous heating power $P_v$ [3]. Furthermore, the electric current density **j** which flows across a metal in steady state satisfies Kirchhoff's principle, i.e. minimization of Joule heating power $P_J$ -see [4][5] and Prob. 3 Sec. 21 of [6].

The aim of this paper is to take advantage of the similarities between the properties of dissipative mechanisms in different physical systems in order to investigate the relaxed states of these systems.

In Sec. 2 we discuss a physical system [7] where Kirchhoff's and Kortweg-Helmholz' principles describe a filamentary structure. In Sec. 3 we show that this structure is a particular realization of a Faddeev-Niemi non-linear sigma model [8]. We discuss the long-wavelength limit of this model in Sec. 4. We show in Sec. 5 that this limit describes relaxation in a class of physical systems much broader than the initial example of Sec. 2. Examples are discussed in Secs. 6, 7 and 8. Conclusions are drawn in Sec. 9.

**2 An application of Kirchhoff's and Kortweg-Helmholz' principles.** Filamentary structures are ubiquitous in plasma physics, both in space [9][10] and in the laboratory [11][7]. In particular, the pinch of a Plasma Focus [12] exhibits remarkably stable filaments [7], which are investigated in [13] with the help of the following results.

Steady configurations of magnetized, quasi-neutral, viscous, electrically conducting fluids which are stable against perturbations of temperature T satisfy the following variational principle [13]:

$$\int dx \left[ \frac{|\mathbf{B}|^2}{2\mu_0} + \frac{\gamma^2}{2} \left| \left( \nabla - i \frac{w}{\gamma} \mathbf{A} \right) \psi \right|^2 - \frac{p_0}{\rho_0} |\psi|^2 + \frac{p_0}{2\rho_0} |\psi|^4 \right] = \min. \quad (2.1)$$

provided that:

i) the fluid is at 'local thermodynamical equilibrium' (LTE) [1] everywhere at all times; LTE means that −although the total system is not at equilibrium− the relationships among thermodynamic quantities (like the internal energy per unit mass, the entropy per unit mass etc.) inside a small mass element of the system are the same as in real thermodynamic equilibrium. LTE is replaced in [14] by i-bis) $Re \ll 1$ and (whenever applicable) i-ter) $Re_m \ll 1$, $Re_m$ magnetic Reynolds number. Assumptions i-bis) and i-ter) mean that inertial effects are negligible in comparison with (resistive or collisional) diffusion effects.
ii) $\nabla T$ is negligible, so that entropy production due to heat conduction and radiation is negligible;
iii) the amount of entropy $dS_p/dt$ produced by chemical reactions and particle diffusion per unit time is negligible. We show in Sec. 5 that this assumption is satisfied e.g. if there is no *net* flux of particles of any chemical species across the boundary and no *net* amount of heat produced by chemical reactions. Together, assumptions ii) and iii) imply that entropy is mainly produced by Joule and viscous heating;
iv) the interaction of the fluid with the external world is negligible in comparison with self-interaction (e.g. magnetic self-interaction energy >> magnetic interaction energy with external electric currents);
v) magnetic helicity $K \equiv \int d\mathbf{x}\, \mathbf{A} \cdot \mathbf{B}$ is not too large [22]. For further discussion of this assumption, see equation (3.2) below.

Here $\mathbf{B} = \nabla \wedge \mathbf{A}$, $\mathbf{A}$, $p_0$ and $\rho_0$ are the magnetic field, the vector potential and the unperturbed values of the pressure p and the mass density ρ of electric charge carriers respectively; moreover, $\mu_0 = 4\pi \cdot 10^{-7}$ T·A$^{-1}$·m, w and γ are constant quantities depending on the particular fluid and $\psi = \psi(\mathbf{x})$ is an order parameter with $\psi^*\psi = \rho$.

It is proven in [15] that if assumptions i)-iii) are satisfied then stability implies minimization of $P_J + P_v$, i.e. simultaneous validity of Kirchoff's and Kortweg-Helmholz' principles. Together with assumption iv), the mass balance, the equation of motion (Navier-Stokes including a Lorenz force density term), Maxwell's equation of electromagnetism and a generalized Ohm's law, such minimization leads to (2.1) in [13].

Formally, (2.1) is similar to the corresponding formula for a type-II superconductor [16] and describes therefore filamentary structures. Moreover, **v** and **B** satisfy:

$$\nabla \wedge \mathbf{v} = r\,\mathbf{v} + w\,\mathbf{B} + \nabla \lambda_1 \quad ; \qquad \nabla \wedge \mathbf{B} = l\,\mathbf{v} + g\,\mathbf{B} + \nabla \lambda_2 \quad ; \qquad (2.2)$$

where r, l and g are further fluid-dependent constant quantities such that $r + g \approx rg - wl \approx |\nabla \wedge \mathbf{v}|^{-1} |\nabla \lambda_1| \approx |\nabla \wedge \mathbf{B}|^{-1} |\nabla \lambda_2| \approx \mathbf{v} \cdot \mathbf{B} \approx |\mathbf{j} \wedge \mathbf{B}| \approx O(\varepsilon)$, $0 < \varepsilon << 1$ and $\lambda_1, \lambda_2$ are harmonic functions of **x** which represent the impact of the external world.

Finally, it has been shown that in the $\varepsilon \to 0$ limit both **v** and **B** are linear superposition of two Beltrami fields with different scaling lengths and are therefore called 'Double Beltrami' vector fields [17]; by definition, a Beltrami field **a(x)** satisfies $\nabla \wedge \mathbf{a} = a\mathbf{a}$ with $a$ pseudoscalar quantity, $\nabla a = 0$ and $1/a$ is the scaling length of **a**.

**3 A non-linear sigma model.** Let our plasma be made of electrons (with electric charge −e and mass m) and ions (with electric charge +e and mass M >> m). Up to terms $\approx$ O(m/M), the R.H.S. of (2.1) coincides with the quantity minimized in equation (1) of [18]. In turn, the latter equation leads to a Faddeev-Niemi non-linear sigma model [8]:

$$F = E_2 + E_4 - V = \min. \quad ; \quad E_2 = \tfrac{1}{2} \int d\mathbf{x}\, (\partial_q \mathbf{n})^2 \quad ; \quad E_4 = \tfrac{1}{2}\alpha \int d\mathbf{x}\, (F_{qu})^2 \quad ;$$

(3.1)

$$F_{qu} \equiv \varepsilon_{ijk}\, n_i\, (\partial_q n_j)(\partial_u n_k) \quad ; \quad i, j, k, q, u = 1,2,3 \quad ;$$

(we invoked both equation (3) of [18] and equations (3)-(8) of [8], where we set $\lambda = 0$ in order to take Coulomb interaction screening into account; see Sec. 5). Here $\alpha$ is a constant quantity, with the dimension of (length)$^2$; $\varepsilon_{ijk}$ is the fully anti-symmetric pseudo-tensor; **n** = **n(x)** is a unit vector (**n**·**n** $\equiv$ 1) which is defined starting from the order parameter, and V is a quantity which does not depend on $\nabla \mathbf{n}$ (unlike $E_2$ and $E_4$). As for different formulations of (3.1), see equation (7) of [18], equation (8) of [8], equation (2) of [19], equation (11) of [20], and equation (2) of [21].

An alternative proof of (3.1) starts from equation (1) of [20] (basically identical to equation (1) of [18]) and leads to equation (11) of [20] (corresponding to (3.1)). According to this proof, our approximation iv) is precisely a necessary condition for validity of (3.1): correspondingly, the contribution of the vector potential **C** due to external sources in [18] is to be neglected. In this proof $\nabla\rho$ too has to be neglected, in agreement with the interpretation of $\psi$ as an order parameter.

According to equation (7) of [8], solutions of (3.1) depend on the value of K, in agreement with our assumption v). A constant $c \propto \alpha^{1/2} > 0$ [23] exists such that [24]:

$$E_2 + E_4 > c\,|K|^{3/4} \qquad (3.2)$$

Far below threshold, the term $\propto$ g in (2.2) is relatively small, and $|\mathbf{j} \wedge \mathbf{v}| \approx O(\varepsilon)$ – see equation (8.2) of [13]. For large-K systems where (3.2) is violated, see [17] and [22].

Finally, according to equation (1.3) of [25], the Euler-Lagrange equation of (3.1) in the Lagrangian coordinate **n** is

$$\mathbf{n} \wedge \Delta \mathbf{n} + (\partial_q F_{qu}) (\partial_u \mathbf{n}) = 0 \qquad (3.3)$$

in the unit system where $\alpha = 1$.

**4 The long wavelength limit.** In spite of their differences, (2.2) and (3.3) refer to the same physical system, and we expect them to agree with each other. In the following we limit ourselves to the long wavelength limit $(\mathbf{n} \cdot \nabla) \mathbf{v} = 0$ (which is relevant to the physics on a spatial scale $\gg \alpha^{\frac{1}{2}}$). In this limit, we show in Appendix that a necessary condition for the agreement of (2.2) and (3.3) is that a pseudoscalar quantity $b$ exists such that $\nabla b = 0$ and that:

$$\mathbf{B} = b\,\mathbf{n} + O(\varepsilon) \qquad (4.1)$$

(Here we are not concerned with the actual value of $b$). Sec. 140 of [26] provides us with an independent confirmation of (4.1), as it shows that if $\mathbf{n}(\mathbf{x})$ varies slowly (again, the long wavelength approximation) and if the reflection $\mathbf{n} \to -\mathbf{n}$ leaves physics unaffected then the most general equations for the unit vector **n** are $\nabla \cdot \mathbf{n} = 0$, $(\mathbf{n} \cdot \nabla)\mathbf{n} = 0$ and $\mathbf{n}\cdot\nabla\wedge\mathbf{n} =$ const. These equations agree with (2.2) and (4.1) up to terms $\approx O(\varepsilon)$.

Equations (2.2) and (4.1) lead to a Double-Beltrami-like structure for **n** and **v**:

$$\nabla \wedge \mathbf{v} = r\,\mathbf{v} + w\,\mathbf{B} + O(\varepsilon) \quad ; \qquad \nabla \wedge \mathbf{n} = l'\,\mathbf{v} + g\,\mathbf{n} + O(\varepsilon) \quad ; \qquad (4.2)$$

where $l' \equiv b^{-1} l$. (By the way, the duality of **n** and **v** is strengthened by the observation that $\mathbf{v} \cdot \mathbf{v} =$ constant in most cases discussed in [13], so that we may take $\mathbf{v} \cdot \mathbf{v} = 1$ with no loss of generality after a suitable change of unit system, in analogy with $\mathbf{n} \cdot \mathbf{n} = 1$). Moreover, (4.1) makes K to reduce to a functional of $\mathbf{n}(\mathbf{x})$, while leaving the well-known connection [27] of K with Gauss linking number unaffected.

The crucial point is that equation (4.2) –the fundamental result of this work– holds regardless of the actual value of $P_J / P_v$. Generally speaking, indeed, (4.2) contains no more explicit information on the original plasma physics problem of [13]. This is equivalent to say that (4.2) holds for a generic, steady-state structure which is stable against perturbations of T, which is described by a unit vector $\mathbf{n}(\mathbf{x})$ –not necessarily connected to an order parameter– which moves across a fluid –not necessarily a plasma– with velocity $\mathbf{v}(\mathbf{x})$, and which satisfies simultaneously:

- either assumption i) or the couple of assumptions i-bis), i-ter);

- assumptions ii), iii), iv);

- assumption v), i.e. inequality (3.2).

Once these conditions are satisfied, steady-state structures which are stable against perturbations of T in different physical systems are obtained from each other simply by changing the relative weight of Joule and viscous heating, i.e. the value of $P_J/P_v$. For example, trefoil solutions of (3.3) exist [19] depending on K, which are observed e.g. in DNA [28]. Usually, indeed, biological fluids satisfy assumption i-bis). Examples of such structures include the filamentary configurations described in Secs. 5, 6 and 7, where $\mathbf{n}(\mathbf{x})$ is the unit vector tangent to the filament at position $\mathbf{x}$. For a problem involving non-filamentary structures, see Sec. 8.

**5 Entropy-raising mechanisms in filaments.** Following the same approach of the last Sections, let us discuss the role of different entropy-raising mechanisms.

As for Joule heating, it occurs only if the filament is electrically conducting. In this case, (4.1) and relationship $|\mathbf{j} \wedge \mathbf{B}| \approx O(\varepsilon)$ in (2.2) imply that charged particles flow along the filament (a physically reasonable picture, indeed) provided that we identify $\mathbf{n}(\mathbf{x})$ with the unit vector tangent to the filament at the point $\mathbf{x}$. Moreover, $\mathbf{j}$ is just an averaged quantity. Then, fluctuations make it still possible to have $P_J = \int d\mathbf{x} <\eta \mathbf{j} \cdot \mathbf{j}> \ > 0$ (where $<>$ and $\eta$ denote averaging and electrical resistivity respectively) even if $\mathbf{j} = 0$ everywhere. Finally, the scaling $|\nabla \wedge \mathbf{B}|^{-1} |\nabla \lambda_2| \approx O(\varepsilon)$ means that magnetic self-interaction of electric currents (if any exists) within the system is much stronger than interaction with external currents. As for electrostatic interactions among different parts of the filament, they are supposed to be screened on the long wavelengths considered here (see Sec. 3).

As for viscous heating, $P_v$ satisfies the condition $P_v(\mathbf{v}) = P_v(-\mathbf{v})$. Accordingly, we may refer to $P_v$ as to the amount heat produced per unit time through viscosity by the motion of the fluid around a filament, as such motion has just velocity $-\mathbf{v}$. Moreover, assumption i-bis) makes the motion to be slow; we take $\nabla \cdot \mathbf{v} = 0$ below. Finally, the scaling $|\nabla \wedge \mathbf{v}|^{-1} |\nabla \lambda_1| \approx O(\varepsilon)$ means that the stresses within the filament depend on the fluid motion far from the filament only weakly.

As for heat conduction and radiation, their role is negligible according to assumption ii).

As for the entropy produced by chemical reactions and particle diffusion, if there are $k = 1,\ldots N$ chemical species then [1] $dS_p/dt = \sum_k \int_\Omega d\mathbf{x} \ T^{-1} \ \mu^0_k \ (dc_k/dt)$, where $\mu^0_k$ and $c_k$ are the chemical potential per unit mass and the mass concentration of the k-th chemical species respectively and the integral is performed on a volume $\Omega$ embedding the filament. Generally speaking, however, we may write $\mu^0_k = \mu^0_k (p, T, c_k)$. If we neglect both $\nabla p$ and $\nabla T$ (which is at least consistent with the above assumption of negligible $\nabla \rho$) and remember that d/dt = $\partial / \partial t$ + **v** $\cdot \nabla$, in steady state we obtain $\mu^0{}_k$(dc$_k$/dt) = (**v** $\cdot \nabla$) g$_k$ = $\nabla \cdot$(g$_k$**v**), where $\nabla \cdot$**v** = 0 and g$_k$ is Gibbs' energy per unit mass of the k-th species. Accordingly, Gauss' theorem of divergence allows us to write TdS$_p$/dt = $\sum_k \int_S$ g$_k$ **v**$\cdot$d**a**. Here integration is performed on the boundary surface $S$ of $\Omega$ with unit surface vector d**a**. Even if exchange of particles and heat with the external world occurs, if there are no net flux of particles of the k-th chemical species and no net amount of heat produced by chemical reactions, then g$_k$ and c$_k$ take the same values everywhere on $S$ and TdS$_p$/dt = ($\sum_k$ g$_k$) ($\int_S$ **v**$\cdot$d**a**) and $\int_S$ **v**$\cdot$d**a** = $\int_\Omega$ d**x** $\nabla \cdot$**v** = 0. Then, we neglect dS$_p$/dt.

Even if assumption i) fails, Kortweg-Helmholz' and Helmholz' principles hold because of assumptions i-bis) and i-ter) respectively. The simultaneous validity of Kirchhoff's and Kortweg-Helmholz' principles which justifies (2.1) means that the filament adjusts its own shape in order to offer both minimum mechanical resistance to the fluid which embeds it and minimum electrical resistance to the electric currents flowing along it. This intuitive conclusion is likely to hold at least in the long wavelength limit, as we discussed the filament structure on scale >> $\alpha^{½}$.

**6 A useful lemma.** Our results do not depend on the actual value of $b$. We are therefore free to choose $b \neq 0$ in the following. We shall derive a result (equation 8) below) which will be useful below. Formally, the fact that **n**$\cdot$**n** $\equiv$ 1 allows us to introduce the dimensionless quantities $\theta$ and $\varphi$ such that **n** = (cos$\varphi$ · sin$\theta$, sin$\varphi$ · sin$\theta$, cos$\theta$) as well as the quantities U, V such that $\varphi$ = − arctan (V/U) and $\theta$ = 2arctan [(U$^2$ + V$^2$)$^{½}$]. Then, it is possible to show [19] that:

$$E_2 = 4 \int d\mathbf{x}\, (1+ U^2 + V^2)^{-2} [(\partial_q U)^2 + (\partial_q V)^2] ;$$

$$E_4 = 16\, \alpha \int d\mathbf{x}\, (1+ U^2 + V^2)^{-4} [(\partial_q U)(\partial_u V) - (\partial_u U)(\partial_q V)]^2 ;$$

(6.1)

The far-away region (corresponding to $\theta \to 0$ in [19]) provides the main contributions to the volume integrals in (6.1). Accordingly, we obtain

$$E_2 \approx 4 \int d\mathbf{x}\, [(\partial_q U)^2 + (\partial_q V)^2] ;$$

$$E_4 \approx 16\, \alpha \int d\mathbf{x}\, (\nabla U \wedge \nabla V)^2 ;$$

(6.2)

Since $b \neq 0$, $E_4 \propto \mathbf{B}_{pol}{}^2$ where $\mathbf{B}_{pol} \equiv \nabla U \wedge \nabla V$ is the generic axisymmetric poloidal magnetic field (U and V play the role of poloidal and toroidal flux respectively). In the long wavelength limit mutual cancellation of small-scale filament wigglings ensures axisymmetry. Ampere's law ensures $\mathbf{B}_{pol} \propto I_{tor} \propto \mathbf{j}\cdot\mathbf{z}$, where $I_{tor}$ and **z** are the toroidal current and the unit vector of the filament axis respectively. In turn, (2.2) and (4.1) give $\mathbf{j}\cdot\mathbf{z} \propto \mathbf{n}\cdot\mathbf{z} \equiv \cos\zeta$ up to terms $\approx O(\varepsilon)$, so that $I_{tor} \propto I \cdot \cos\zeta$ and

$$E_4 \propto I^2 \cdot (\cos\zeta)^2 \qquad (6.3)$$

**7 Double helix.** Let us compute F for a system of two filaments I and II, which are the mirror image of each other. Since K is a pseudoscalar then $K_I + K_{II} = 0$. We consider two cases: a) I and II are both straight ($\zeta = 0$) and are at large distance from each other; b) th axes of I and II coincide and I, II form a double helix ($\zeta > 0$).

**Case a).** In this case we are free to take $V = 0$. Since I and II are at large distance from each other, we compute $F_I$ ($F_2$) just in absence of II (I). Then $F_{case\ a} = F_I + F_{II}$. Analogously, $K_H = K_I + K_{II} = 0$. Since I and II are the mirror image of each other and F is a true scalar quantity, $F_I = F_{II}$. Then $F_{case\ a} = 2\ F_I$. But [19] $F_I = E_{2\ I} + E_{4\ I}$, and the same holds for $F_{II}$; moreover, once applied to a single filament I (II) the virial theorem ensures that $E_{2\ I\ (II)} = E_{4\ I\ (II)}$. Accordingly, $F_{case\ a} = 4E_{4\ I}$. Moreover, (6.3) gives $E_{4I} \propto I_I^2 \cdot (\cos\zeta)^2 = I_I^2$ for $\zeta = 0$, and $F_{case\ a} = 4\ I_I^2$.

**Case b).** The double helix is the mirror image of itself. Then the pseudoscalar quantity K still vanishes. Transition from a) to b) preserves K, even if it affects $\zeta$. Generally speaking, however, V vanishes no more. Let us write $V = -E_0$, so that $E_0 > 0$ corresponds to attraction between filaments. We have still to compute $E_2$ and $E_4$.

Let us denote with $I_{I\ (II)}$ the current flowing across I (II). Since I and II are the mirror image of each other, $I_I = I_{II}$. Moreover, $\cos\zeta$ is the same in I and II, and the same result holds for all proportionality constants. The current which flows across the double helix is $I_I + I_{II} = 2\ I_I$. Then $E_4 \propto (I_I + I_{II})^2 \cdot (\cos\zeta)^2$, i.e. $E_4 = 4\ E_{4I} \cdot (\cos\zeta)^2 = F_{case\ a} \cdot (\cos\zeta)^2$. As for $E_2$, the virial theorem holds no more as $V \neq 0$; all the same, we may still write $E_2 > 0$. Let us define $\Gamma \equiv 1 + E_2/E_4$ (with $\Gamma > 1$). Formally, we write: $F_{case\ b} = E_2 + E_4 - E_0$, i.e.

$F_{case\ b} = \Gamma \cdot F_{case\ a} \cdot (\cos\zeta)^2 - E_0$

The double helix is stable in comparison with the separate filaments whenever $F_{case\ b} < F_{case\ a}$, i.e.

$(\Gamma - 1) \cdot (\cos \zeta)^2 < E_0/\ F_{case\ a}$ (7.1)

Since $F > 0$ in all cases and $\Gamma > 1$, according to (7.1) double helix stability requires that $E_0 > 0$, i.e. that filaments attract each other, as expected. (In the long wavelength limit (3.2) may still be satisfied even for $K \neq 0$ as $c \propto \alpha^{1/2} <<$ the typical spatial scale of the system). Moreover, we may destabilize the double helix by increasing $(\cos\zeta)^2$, i.e. by making filaments straighter ($\zeta \rightarrow 0$). Finally, straightforward computation shows that triple helices are even more difficult to stabilize.

**8 Skyrmion writing and Joule heating.** Minimisation of $E_2 + E_4$ in (3.1) is often invoked when describing skyrmions, which are topologically stable spin-swirling objects [29] recently utilised in condensed matter physics where a unit vector **n**(**x**) is defined. Recently, it has been reported [30] that individual skyrmions can be written and deleted in a controlled fashion with local spin-polarized currents applied to magnetic films. Now, currents dissipate. Accordingly, this 'skyrmion writing' is a far-from-equilibrium, entropy-raising, hopefully reproducible (i.e., stable) process. It is meaningful to ask whether our analysis may provide us with some relevant information.

Let us focus our attention on a beam of polarized electrons with current $I_b$. A polarized electron with spin **s** and momentum **p** carries an amount $\propto$ **s** · **p** of magnetic helicity K, and it is the total amount of the latter which rules the structure of the solution of (3.1). Once it has entered the magnetic film, the system made of the beam and the surrounding, electrically conducting medium undergoes Joule heating, and (after a suitable transient) attains a state described by Kirchhoff's principle (no viscous heating occurs). Thus, we deal with skyrmion writing as with a problem of electric conduction. Our macroscopic approach drops all detailed microscopic physics including spin, etc. and is able to provide us just with a scaling law involving macroscopic quantities, as we shall see below.

Polarisation of electrons allows the beam to feed the film with K, and the film may therefore form skyrmions. Externally applied magnetic fields [30] tune the energy landscape, and the temperature is adjusted to prevent thermally activated switching between topologically distinct states. As far as the energy W of the skyrmion is > T, thermal bath effects are negligible. But W is supplied by the beam itself, and is therefore proportional to $I_b^2$: say, $W = \kappa\, I_b^2$.

If the value of K in an isolated skyrmion is $K_s$, then (3.2) implies $F > c\, |K_s|^{3/4}$ where $c$ depends on the material. Moreover, virial theorem for an isolated skyrmion ensures that $F = 2\, E_4$. Furthermore, we may reasonably write $E_4 = \xi\, W$ (where we identify W with the magnetic energy, take $\xi$ = constant, and choose $b \neq 0$ as usual). These relationships give a lower bound on the beam current which induces reproducible formation of a skyrmion:

$$I_b > (c / 2\, \xi\, \kappa)^{½}\, |K_s|^{3/8} \quad\quad\quad (8.1)$$

As a matter of principle, (8.1) is an experimentally verifiable prediction.

**9 Conclusions.** We show that a Faddeev-Niemi non-linear sigma model [8] describes in the long wavelength limit a wide class of steady-state, knotted configurations of physical systems far from thermodynamical equilibrium which are stable against perturbations of temperature and interact weakly with the external world, provided that temperature gradients are negligible, the assumption of 'local thermodynamical equilibrium' (LTE) holds, entropy is mainly produced through Joule and viscous heating, and the Gauss linking number is lower than a threshold. (LTE may be replaced by the assumption that inertial effects are negligible in comparison with diffusion effects).

These configurations include a broad class of filamentary structures routinely observed in plasma physics [7]. However, they seem to be relevant to many other fields, as spontaneous relaxation to knotted configurations turns out to be e.g. the final outcome of the evolution of waves propagating across dispersive media described by non-linear Schroedinger equation [31].

Regardless of their detailed microscopic structure, in the long wavelength limit stable filaments adjust themselves in order to offer minimum resistance to the medium (fluid, plasma) embedding them and to the electric currents (if any) flowing across them. Thus, irreversibility affects the very topology of the final outcome of a relaxation process − for other examples, see [2]. Both the velocity and the unit vector tangent to the filament are approximately described as Double Beltrami structures [17]. The Gauss linking number of the filament is related to the magnetic helicity whenever electric currents flow along the filament.

As a particular example, we derive a necessary condition for stability of a double helix, which turns out to be critically dependent on the twistedness of each helix even if an attractive potential acts between the helices.

Moreover, our discussion of entropy production in the stable configurations quoted above allows us to we provide a simple model of the recently discovered writing procedure of a skyrmion in a magnetic film with the help of a beam of polarised electrons. It turns out that a minimum beam current is required in order to write a single skyrmion.

**Acknowledgments.** Useful discussions with Prof. W. Pecorella, Univ. Tor Vergata, Roma are gratefully acknowledged.

**Appendix. Proof of (4.1).** Let us define $\mathbf{Z} \equiv \mathbf{B} - b\,\mathbf{n}$. To start with, we note that $F_{qu}$ is an antisymmetric tensor and that $\mathbf{v}$ and $\mathbf{B}$ are the only vectors relevant to (2.2) up to terms $\approx O(\varepsilon)$, so we may take

$$F_{qu} = \varepsilon_{qus}\,(f\,v_s + h\,B_s) + O(\varepsilon) \tag{A.1}$$

where $s = 1,2,3$, $v_s$ and $B_s$ are the s-th components of $\mathbf{v}$ and $\mathbf{B}$, and $f$, $h$ are constant quantities. If (2.2) agrees with (3.3), then these equations must allow unambiguous computation of a and c up to terms $\approx O(\varepsilon)$. We show that (4.1) is just a consequence of this requirement. Firstly, relationships $\mathbf{n}\cdot\mathbf{n} \equiv 1$, $|\mathbf{j} \wedge \mathbf{v}| \approx O(\varepsilon)$, Ampère's law $\nabla\wedge\mathbf{B} = \mu_0\mathbf{j}$ and Gauss' law $\nabla \cdot \mathbf{B} = 0$ lead to:

$$\nabla(\mathbf{B}\cdot\mathbf{B}) = O(\varepsilon) + O(|\mathbf{Z}|)\ ;\ |\mathbf{v}\wedge\nabla\wedge\mathbf{B}| = O(\varepsilon)\ ;\ \nabla\cdot\mathbf{n} = O(\varepsilon) + O(|\mathbf{Z}|) \tag{A.2}$$

Finally, in the long wavelength limit (2.2), (3.3), (A.1), (A.2) and $\mathbf{v}\cdot\mathbf{B} \approx O(\varepsilon)$ lead to:

$$X\,(\mathbf{B} \wedge \nabla \wedge \mathbf{B}) + Y\,(\mathbf{B} \wedge \nabla \wedge \mathbf{v}) = O(\varepsilon) + O(|\mathbf{Z}|) \tag{A.3}$$

where $X \equiv b^{-1}\,(f\,w + h\,g) - b^{-2}\,g$ and $Y \equiv b^{-1}\,(f\,r + h\,l) - b^{-2}\,l$. In turn, (A.3) holds as an identity for arbitrary $|\mathbf{v}|$ and $|\mathbf{B}|$ provided that

$$X = O(\varepsilon) + O(|\mathbf{Z}|)\ ;\ Y = O(\varepsilon) + O(|\mathbf{Z}|) \tag{A.4}$$

Toghether with $r + g \approx O(\varepsilon)$ and $r\,g - w\,l \approx O(\varepsilon)$, (A.4) allows us to compute both $f$ and $h$ up to terms $\approx O(\varepsilon)$ provided that both $b$ and the 4 quantities r, g, l and w (i.e. $\mathbf{v}$ and $\mathbf{B}$ in the $\varepsilon \to 0$ limit) are known, and provided that $O(|\mathbf{Z}|) \approx O(\varepsilon)$, i.e. that (4.1) holds.